\documentstyle[epsf,psfig,12pt]{article}
\newcommand{\nin}{\noindent}

\newcommand{\be}{\begin{equation}}
\newcommand{\ee}{\end{equation}}
\newcommand{\bea}{\begin{eqnarray}}
\newcommand{\eea}{\end{eqnarray}}

\begin{document}
\begin{center}
\vspace{8mm}

{\bf DIRECTIONAL EMISSION FROM ASYMMETRIC RESONANT CAVITIES}
\vspace{8mm}

{Jens U.~N{\"o}ckel, A.~Douglas Stone, Gang Chen, 
Helene L.~Grossman and Richard K.~Chang
\\}
{\footnotesize{\it Department of Applied Physics and Physics, Yale University
\\P.~O.~Box 8284, New Haven Connecticut 06520-8284\\}}
\vspace{10mm}
{{\em Published in} Optics Letters {\bf 21}, 1609 (1996)}\\
\vspace{10mm}

{ABSTRACT}\\
\parbox[t]{30pc}{
\nin
{\footnotesize
Asymmetric resonant cavities (ARCs) with highly non-circular but
convex cross-sections are predicted theoretically to have 
high-Q whispering gallery modes with very anisotropic emission.
We develop a ray dynamics model for the emission pattern and
present numerical and experimental confirmation of the theory.
}
}
\end{center}

\vspace{10mm}
In recent work \cite{wgopt,prl,chapter}, a new class of optical 
resonators has been proposed, comprised of convex dielectric 
bodies which are substantially deformed from spherical or
cylindrical symmetry.  We have shown \cite{wgopt,chapter} that such
asymmetric resonant cavities (ARCs) still retain high-$Q$ ($Q > 1000$) 
whispering
gallery (WG) modes up to distortions as large as 50\% of the undeformed
radius R; at the same time ray simulations indicated that the emission 
pattern from 
these modes becomes highly anisotropic \cite{wgopt}.  The initial work
focused on cylindrical ARCs; subsequently \cite{prl} 
we generalized the ray dynamics model to explain the experimentally
observed directional lasing emission from non-spherical microdroplets. 
In this work we propose for the first time a general framework for predicting
the high-emission directions for ARCs.  
Although generically the ray dynamics 
in ARCs is chaotic, by using the Poincar\'e surface of section method and
the adiabatic approximation, it is possible to find universal features in 
the ray motion which determine the high emission points.
Interestingly, the latter do not always coincide with the regions of 
highest curvature as might be expected.
We compare the ray model to the
solutions of the wave equation for cylindrical ARCs and to the experimentally
determined lasing emission pattern from deformed liquid columns and find
good agreement.

The WG modes of {\it undeformed} spheres or cylinders have high $Q$
because the rays impinge on the boundary with a conserved angle of incidence 
$\sin \chi > \sin \chi_c\equiv 1/n$, ($n$ is the refractive index) 
and are thus trapped by total internal reflection.
The intrinsic width of these resonances then arises only due to
evanescent leakage.
In contrast, in ARCs $\sin \chi$ is not conserved and the dominant
escape mechanism is refractive escape: starting from a WG orbit 
with $\sin \chi > \sin \chi_c$, after a large number of reflections 
the ray may impinge on
the boundary below the critical angle and escape with high probability.
The high intensity points in the near-field correspond to the regions where
most of the refractive escape occurs; the far-field directionality
must be determined by following the bundle of refracted rays.

In order to understand the high emission directions from ARCs, it is essential
to analyze the ray dynamics in phase space, and not just by ray-tracing in
real space.  That is because the partially chaotic (mixed) phase space
exhibits remnants of regular structure\cite{wgopt,prl,chapter},
creating dominant flow patterns which
determine the high intensity points. The standard technique used 
in non-linear dynamics to obtain an understanding of a mixed phase
space is the Poincar{\'e} surface of section (SOS). It is obtained
by plotting, for successive reflections of a ray, the angular position
$\phi$ (azimuthal angle for cylinders, polar angle for spheroids) along 
the boundary where the reflection occurs, and the value of $\sin\chi$.
Recording on the SOS a relatively small number of ray trajectories 
($\sim 10-20$) for about $500$ reflections 
yields a detailed picture of the phase space structure. It typically 
exhibits ``islands'' and ``random'' disconnected points, associated
with stable and chaotic trajectories, respectively
\cite{wgopt,chapter}.  Certain properties of the
SOS are generic for all ARCs and can be used to deduce general features of 
the escape directionality.

These generic features are illustrated by the ray dynamics of 
a cylindrical ARC with a 2D quadrupolar deformation of its cross-section
(normalized to constant area) given in cylindrical coordinates by 
\be
r(\phi)=\frac{1}{\sqrt{1+\epsilon^2/2}}\,(1+\epsilon\,\cos 2\phi).
\ee
The parameter $\epsilon$ measures the degree of
deformation, the aspect ratio being 
$(1 - \epsilon)/(1 + \epsilon) \approx 1 - 2 \epsilon$. For a circle, 
$\epsilon=0$; when $\epsilon\ge 0.2$ the shape becomes 
non-convex and no WG caustics 
remain in existence \cite{mather}.  The interesting regime for this
shape deformation then lies in the interval $0.05<\epsilon<0.15$.
Equation (1) describes the simplest 2D
ARC because any dipolar component can be removed by a 
shift of origin,
and all other ARCs will involve higher multi-poles.  For example, the
ellipse (which is the only 2D ARC which has no chaotic trajectories)
has multipoles of all even orders although it agrees with
the quadrupole to order $\epsilon$. 

The partial SOS shown in Fig.~1 is generated for $\epsilon=0.072$ for which
WG trajectories with $\sin \chi< 0.8$ are no longer
confined by a caustic and are chaotic; ultimately such orbits will diffuse 
to lower $\sin \chi$, reach $\sin \chi_c$ and escape refractively.
However, if we follow these trajectories for 200 reflections as is done
in Fig.~1, they are seen to fall close to curves in the SOS which are
given by\cite{robnik}
\be\label{adiab}
\sin\chi(\phi)=\sqrt{1-\left(1-S^2\right)\kappa(\phi)^{2/3}},
\ee
where $\kappa(\phi)$ is the curvature of the interface and $S$ 
is a constant parametrizing the curves, roughly equal to
the average value of $\sin \chi$.
The relation (2) represents an {\it adiabatic} approximation valid for 
$d\kappa/d\phi\sqrt{1-\sin^2\chi}\ll 1$, which describes the ray 
dynamics for intermediate times rather well, as shown in  Fig.~1.
It does not describe the {\it chaotic} behavior of the rays 
and hence fails badly
at long times when the ray will leave the adiabatic curve.
\begin{figure}[bt]
\epsfbox[0 60 100 250]{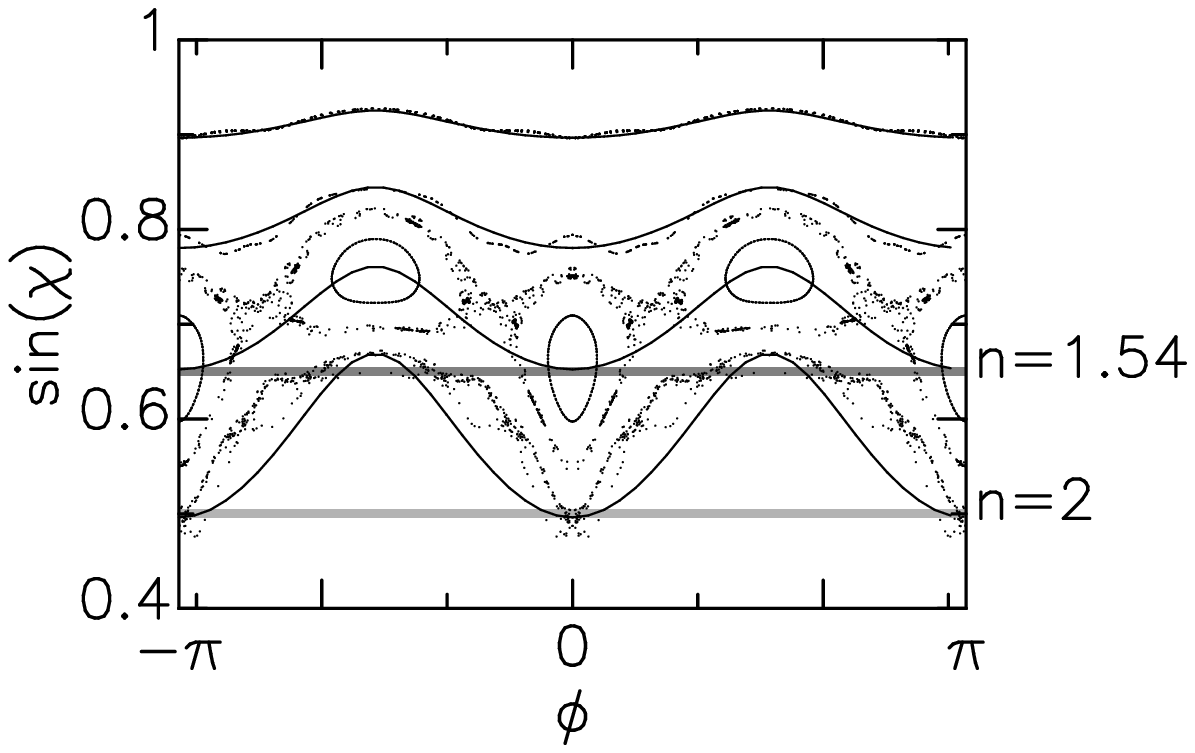}
\vspace{4mm}
\footnotesize\noindent
Figure 1: SOS for quadrupole at $\epsilon=.072$, showing four chaotic
trajectories followed for 100 - 200 reflections. Superimposed are the
curves given by Eq.~(2). Shaded horizontal lines indicate $\sin\chi_c$
for different refractive indices $n$.
\label{fig1}
\end{figure}\normalsize

While the failure of the adiabatic approximation for long times must be 
taken into account in calculating the {\it broadening} of the resonance due to
refractive escape, the approximation is adequate for determining the 
resonance {\it frequency} using eikonal theory \cite{later}.
In this approach (which neglects broadening) the
adiabatic constant $S$ in Eq.~(\ref{adiab}) can take on only a
discrete set of values $S_{pm}$, where the integers $p$ and $m$ reduce
to the radial and azimuthal mode indices, respectively, of the
state for circular cross-section.  For any deformation $\epsilon$
and for any resonance $(p,m)$  we can find the 
adiabatic invariant curve (AIC) parametrized by $S_{pm}$ along which 
the corresponding rays move.  If this AIC intersects the 
critical line $\sin\chi_c$, then such resonances are very
short-lived and may be ignored. Labeling by $S_c$ the
AIC whose minima are tangent to the critical line, the the long-lived WG
modes are characterized by $S_{pm}>S_c$. 
A chaotic trajectory starting at $S=S_{pm}$ eventually deviates from this
curve by diffusion with a bias towards smaller $S$ ($\sin \chi$).
At each step of this diffusion process a ray moves for intermediate times along
a fixed AIC.  Ultimately the ray
reaches the critical AIC with $S=S_c$ which is tangent to $\sin\chi_c$ and 
the ray rapidly flows along this AIC to the minima
where it escapes refractively with high probability.
Therefore assuming that the adiabatic approximation is valid for intermediate
times, escape will occur only at or near the tangency points of $S_c$.
[e.g. $\phi=0,\pi$ For the quadrupole billiard (Fig.~1)].
From Eq. (2) we see that quite generally the minima of the AIC's
occur at the point(s) of maximum curvature.  Hence this argument suggests
that escape occurs primarily near the points of maximum curvature, as
might be intuitively expected.
Furthermore, because most rays escape from the tangent AIC,
they just barely violate the total internal reflection condition and are
emitted almost tangent to the boundary.  The rays are therefore 
{\it not} strongly dispersed in angle by refraction, and the emission pattern
is peaked in both the near-field {\it and} far-field, with the far-field
emission peaks simply rotated by $\pi/2$.
\begin{figure}[bt]
\epsfbox[0 60 100 300]{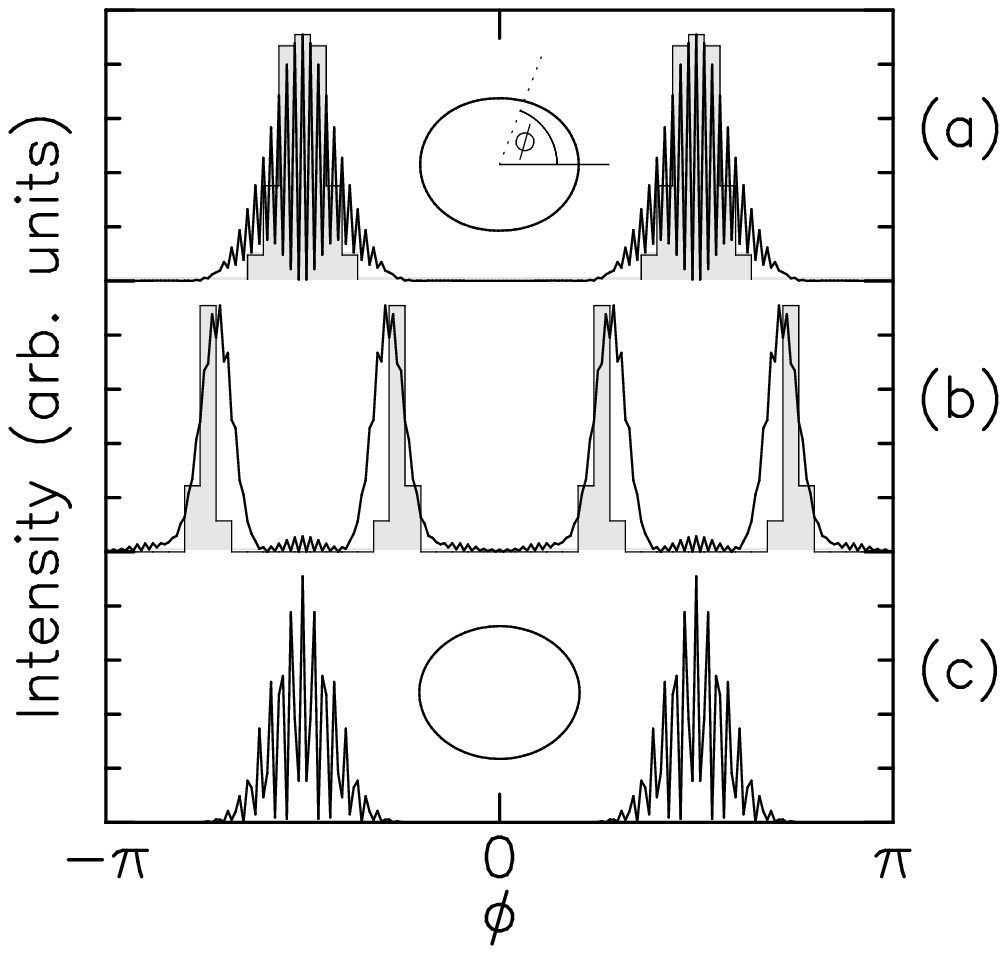}
\vspace{4mm}

\noindent\footnotesize
Figure Far-field emission directionality from quadrupole (a,b) and
ellipse (c) with distortion $\epsilon=0.09$, 
for a resonance $m=68$. Solid
lines: intensity from solution of the wave equation; histograms:
classical 
prediction. The refractive index is $n=2$ in (a) and $n=1.54$ in
(b,c). The size parameter at this $\epsilon$ is $kR=45.2$ in (a), 
$kR=48.1$ in (b) and $kR=48.0$ in (c). The insets show the shapes.
\label{fig2}
\end{figure}

\normalsize
These conclusions are independent of the particular starting
curve $S_{pm}$ as long as it is larger than $S_c$.  Hence we 
reach the central conclusion of this work: there exists a 
{\em universal} and highly directional emission pattern for all
WG resonances of cylindrical (2D) ARCs corresponding to $S_{pm}>S_c$.
The actual pattern can be generated approximately 
by simply following a ray ensemble begun
on the AIC with $S=S_{pm}$ and making a histogram of the number of rays
escaping at a given angular position (near-field) and, after refraction,
escaping in a given direction (far-field) \cite{prl}.  We refer to these
as the ``classical'' directionality histograms.

The resulting far-field directionality is shown in Fig.~2 (a) in 
comparison with the intensity pattern obtained by exact numerical solution for
the quasi-bound state.  The good agreement shown is also found
for all other resonances studied at this index of refraction.
The wave results are obtained by solving for the
quasibound states\cite{young} of a deformed cylinder assuming TM
polarization. Our algorithm makes use of the Rayleigh
hypothesis\cite{millar} to expand internal and external fields in Bessel
functions, and then imposes appropriate matching conditions at the interface, 
satisfied for complex wavevector.

Experimental evidence for the universality of the emission directionality can
be obtained from a measurement of the lasing emission
produced by liquid dye columns. In a variant on the well-studied lasing
emission from micro-droplets\cite{prl}, ethanol containing Rhodamine B dye
was forced through circular and rectangular orifices without the usual 
segmentation of the stream.  The dye column produced by the circular orifice 
(of radius $75 \mu \rm m$) is cylindrical, whereas that
produced by the rectangular orifice (of dimension $1000 \mu \rm m 
\times 25 \mu \rm m$) has an oval cross-section with an 
eccentricity which decays (because of viscous damping) until it is nearly 
circular at $2\,$cm from the orifice.  The cross-sectional distortion
which experiences the smallest damping is quadrupolar, so that we
expect the dominant deformation to be roughly given by Eq.~(1). 
Hence the dye column at the appropriate height provides a realization
of a cylindrical ARC.  The surface tension causes the 
major axis of the quadrupole to oscillate in orientation with
respect to the long axis of the orifice.

From previous studies of droplets the lasing of the column should involve
many WG modes; however if our universality hypothesis is correct we should
still observe highly directional emission from the deformed jet since all
such WG modes have the emission pattern of Fig.~2(a).  Specifically, we expect
to see high emission intensity in the direction 
perpendicular
to the long axis of the deformed cross section which will be rotated
by $90^{\circ}$ with each half-cycle of the quadrupolar oscillation.

\begin{figure}[bt]
\vspace*{-4cm}
\hbox{\hspace*{0.5cm}\psfig{figure=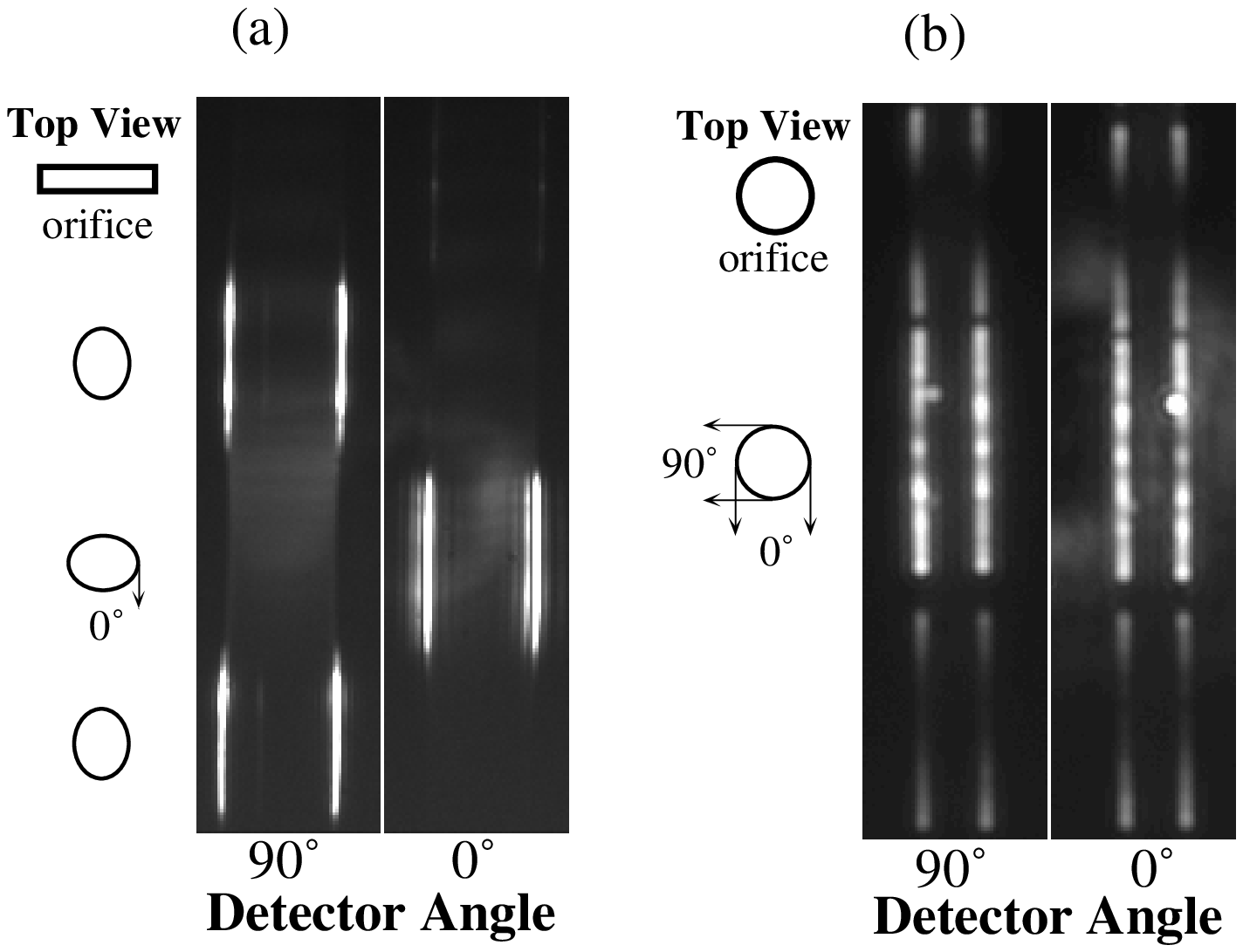,width=15cm}}

\vspace*{-7.5cm}
\noindent\footnotesize
Lasing emission from a liquid dye column created by rectangular (a)
and circular (b) orifices. Simultaneous images taken at $0^{\circ}$
and $90^{\circ}$ angle with respect to the pump beam are shown side by
side. 
\end{figure}
The dye column was pumped with a $537\,{\rm nm}$ pulsed dye laser
with pulse duration $\approx 5 {\rm ns}$.  The pump laser is oriented 
perpendicular to the long-axis of the rectangular orifice.  Two
collection lenses
set to f/22 were placed at $0^{\circ}$ and $90^{\circ}$ relative to the pump 
laser beam.  By using mirrors and a beam splitter, lasing images
produced by the two lenses were combined side by side on a single 
CCD detector and were
recorded simultaneously with appropriate calibration to preserve spatial
correspondence anywhere along the column.  As predicted by the theory,
we observe [Fig.\ 3(a)] a striking oscillatory rotation of the high emission 
intensity between
the $90^{\circ}$ and $0^{\circ}$ images, commensurate with the oscillation
of the deformation of the cross section.  No such oscillatory behavior is
observed in the lasing emission from the circular liquid column 
[Fig.\ 3(b)]. 

The prediction of emission from the points of maximum $\kappa$ is based
on Eq.~(\ref{adiab}). However, the actual phase-space flow pattern departs
strongly from that predicted by the adiabatic approximation when the AIC
intersects large islands of stable motion as can be clearly seen in the SOS
of Fig.\ 1.  If the index of refraction is such that 
$\sin\chi_c$ intersects these large islands at the minima of
the AIC, then the relevant trajectories must circulate around these
islands.  Hence rays escaping at the critical line
do not do so at the points of maximum curvature but at two points
offset by roughly the width of the island.  The resulting near-field 
pattern is complex due to interference of these two ray bundles, but
often shows a {\it dip} at the points of maximum curvature.
Moreover, in all cases the far-field pattern shows
strong splitting of the high emission peaks [Fig.~2(b)] in
contrast to the two peaks of Fig.~2(a) for both the ray histograms and
the exact wave solutions.
We refer to this remarkable phenomenon as {\em dynamical eclipsing}.
It deserves further experimental and theoretical study, however the
refractive index of $n \approx 1.5$ required for this is not
conveniently reached with liquids.

It should be emphasized that the only
parameter we changed in order to find the peak-splitting in our 
calculations is the index of refraction and that the splitting
occurs for all resonances we have tested at this index \cite{later}.
To confirm further that the numerically determined peak-splitting actually
has the classical (ray dynamical) origin we propose, we calculate in 
Fig.~2(c) the resonances of an elliptical ARC at the same index and
aspect ratio as the quadrupole of Fig.~2(b).  
The ellipse generates no chaos and indeed the adiabatic approximation
of Eq.~(2) becomes exact.  Hence there are no 
islands near the critical line in the SOS, and
we find no peak splitting despite
the high similarity between the two shapes in Fig.~2(b),(c).
Thus we believe that this phenomenon provides an unambiguous and striking
fingerprint of the classical phase space structure in 
the emission directionality. 

We acknowledge partial support by U.~S.~Army Research Office 
Grant DAAH04-94-G-0031 and NSF Grant DMR-9215065 and thank
Ying Wu and Marko Robnik for helpful discussions.

\end{document}